\documentclass[useAMS,usenatbib,usegraphicx,referee]{mn2e}
\usepackage{natbib}
\usepackage{lscape}
\usepackage{longtable}
\usepackage{Times}

\newcommand{\msun}{M_\odot}
\newcommand{\lsun}{L_\odot}
\newcommand{\kpc}{\rm{kpc}}
\newcommand{\pc}{\rm{pc}}
\newcommand{\gyr}{\rm{Gyr}}

\newcommand{\rhlcut}{r^{\rm{cut}}_{hl, \rm{obs}}}
\newcommand{\rcncut}{r^{N, \rm{cut}}_{c, \rm{obs}}}
\newcommand{\rclcut}{r^{L, \rm{cut}}_{c, \rm{obs}}}
\newcommand{\rcobs}{r_{c, \rm{obs}}}
\newcommand{\henon}{H\'enon}
\newcommand{\rcn}{r^{N}_{c, \rm{obs}}}
\newcommand{\rcl}{r^{L}_{c, \rm{obs}}}
\newcommand{\rhl}{r_{hl}}
\newcommand{\moverl}{M_{\rm{cl}}/L_{\rm{cl}}}
\newcommand{\moverlsun}{M_\odot/L_\odot}
\newcommand\aj{{AJ}}
\newcommand\apj{{ApJ}}
\newcommand\apjl{{ApJ}}
\newcommand\apjs{{ApJS}}
\newcommand\aap{{A\&A}}
\newcommand\mnras{{MNRAS}}
\newcommand\pasp{{PASP}}

\title[Understanding Core-Collapsed Clusters]
  {Understanding the Dynamical State of Globular Clusters: Core-Collapsed vs Non Core-Collapsed}
\author[S. Chatterjee et al.]
  {Sourav~Chatterjee,$^1$\thanks{email:s.chatterjee@astro.ufl.edu} 
  Stefan~Umbreit,$^2$
  John~M.~Fregeau,$^3$
  and Frederic~A.~Rasio$^{2}$\\
  $^1$Department of Astronomy, University of Florida, Gainesville, FL 32611, USA. \\
  $^2$Center for Interdisciplinary Exploration and Research in Astrophysics (CIERA) and \\
  Department of Physics \& Astronomy, Northwestern University, 2145 Sheridan Road, Evanston, IL 60208, USA\\
  $^3$Institute for Defense Analyses, 4850 Mark Center Dr, Alexandria, VA 22311.  }

\pagerange{\pageref{firstpage}--\pageref{lastpage}} \pubyear{2002}

\def\LaTeX{L\kern-.36em\raise.3ex\hbox{a}\kern-.15em
    T\kern-.1667em\lower.7ex\hbox{E}\kern-.125emX}

\begin{document}

\label{firstpage}

\maketitle

\begin{abstract}
 We study the dynamical evolution of globular clusters using our H\'enon-type Monte Carlo 
code for stellar dynamics including all relevant physics such as two-body
relaxation, single and binary stellar evolution, Galactic tidal stripping, and
strong interactions such as physical collisions and binary mediated
scattering.  We compute a large database of several hundred models starting
from broad ranges of initial conditions guided by observations of young and
massive star clusters. We show that these initial conditions very naturally
lead to present day clusters with properties including the central density,
core radius, half-light radius, half-mass relaxation time, and cluster mass, that match well with those of
the old Galactic globular clusters. In particular, we can naturally reproduce
the bimodal distribution in observed core radii separating the
``core-collapsed" vs the ``non core-collapsed" clusters. We see that the
core-collapsed clusters are those that have reached or are about to reach the
equilibrium ``binary burning" phase.  The non core-collapsed clusters are still
undergoing gravo-thermal contraction.  
\end{abstract}


\section{Introduction}
Studying the evolution of dense star clusters, such as old globular clusters (GCs),
is of great interest for a variety of branches of astronomy and astrophysics.
The high central densities and high masses of GCs
make them hotbeds for strong dynamical interactions facilitating formation 
of many exotic sources (e.g., X-ray binaries, millisecond radio pulsars,
type Ia supernovae, and blue straggler stars).
GCs are important targets for extragalactic Astronomy. Detailed observations of their spatial
distribution in a galaxy can constrain, for example, the potential of
the dark matter halo, and give clues to the assembly history of the galaxy. The old ages of 
GCs provide a direct window into the major star-formation episodes in the early universe.  

One long-standing question regarding GCs concerns the nature of their progenitors.  
The observed young massive star clusters 
\citep[e.g.,][]{1992AJ....103..691H,1995AJ....109..960W,1997AJ....114.2381M,2007A&A...469..925S,2009gcgg.book..103S} 
seem to be potential 
progenitors of the current GCs.  The masses, typical sizes, and inferred dissolution timescales for the 
so-called ``super star clusters", as observed, for example, in M51 \citep[][]{2007A&A...469..925S}, make 
them especially good candidates.  Interestingly, similar to the Galactic GCs (GGC), the 
sizes of observed super star clusters are typically a few parsecs independent of the cluster mass \citep[e.g.,][]{2007A&A...469..925S,2009gcgg.book..103S}.                
However, the link between these super star clusters
and the old evolved GGCs as well as GCs in other galaxies remains speculative.  

The main difficulty in establishing a link between the two populations is that almost a Hubble time of 
evolution separates them. Although the individual qualitative effects 
of each physical process in a GC has been known 
from decades of numerical studies \citep[e.g.,][]{2003gmbp.book.....H}, it is impossible to 
estimate the collective 
effect of these interdependent processes unless a self-consistent simulation is done 
including them all.  
For example, two-body relaxation leads to 
a slow energy diffusion from the core to the halo in a GC leading to a slow and steady contraction 
of the core until the gravo-thermal instability occurs and the core collapses under gravity.  
Since, due to equipartition of energy, the low-mass stars are scattered to higher velocities relative 
to their high-mass counterparts, two-body relaxation also leads to mass segregation in the 
cluster.  The high-mass stars sink to the core and the low-mass stars escape to the halo and 
preferentially get stripped through the tidal boundary of the cluster.  Because of this preferential loss 
of low-mass stars, the stellar MF changes with time.  The stellar MF, in turn, determines
the fraction of low and high-mass stars in a 
cluster as well as the average stellar mass, affecting the mass-segregation timescale
\citep{2004ApJ...604..632G}.  
The contraction of the core via two-body relaxation increases its density. The core density
determines the interaction rate in the core.  These rates affect the binary-single (BS) and binary-binary (BB) interaction 
probabilities at a given time.  The BS and BB interactions in turn generate energy in the core and 
can support the core stopping further contraction, thus, affecting the central density. The BS and 
BB interactions also change the orbital properties of the binaries taking part in these scattering 
interactions.  These changes in the binary orbits can alter the evolution pathways 
that would be taken by a given binary which 
in turn changes, for example, the rate of formation of exotic stellar populations such as 
X-ray binaries and blue straggles stars (BSS).    
Due to this complexity in the evolution of dense star clusters, the only way to learn more about the 
possible initial properties of the observed GCs is to perform numerical simulations including 
all of the above physical processes in tandem with reasonable accuracy for a large enough $N$. 

The study of GCs has a somewhat long and varied history during which numerical simulations 
and observations have complemented each other \citep{2003gmbp.book.....H}.  
In particular, understanding the physical processes in the cores 
of GGCs has been of prime interest since the evolution of these dense clusters are 
driven mainly by the energy generation in the core and the transport of this energy from the core.  
The GGCs are observed to show a clear bimodal distribution of the core sizes that separates the so-called 
``core-collapsed" clusters from the non core-collapsed clusters \citep{1996AJ....112.1487H,2005ApJS..161..304M}.  
The core-collapsed clusters in general show a power-law slope in their density profiles near the center.  
In contrast, the density profiles of the non core-collapsed clusters are well described by a King profile 
\citep{1966AJ.....71...64K} and show a clear flat part near the center.  
Theoretical analysis based on the current estimated relaxation times for the GGCs indicate that the 
majority of the GGCs should have had a deep collapse \citep{1993ASPC...50..373D,1996AJ....112.1487H}.  
Thus, before it was found that all GGCs contain 
dynamically significant numbers of binaries, theoretical studies focused on understanding the process 
of core-collapse via the balance of outward diffusion of energy from the core due to two-body relaxation and post-collapse 
evolution due to dynamical formation of binaries \citep[e.g.,][]{2003gmbp.book.....H}.  After it was observed in the early 1990s 
that all GGCs contain sufficient number of binaries such that they must have been born with substantial 
primordial binary populations, theoretical studies focused on properties of clusters in the ``binary-burning" 
phase in which the outward diffusion of energy via two-body relaxation is balanced by 
production of energy via dynamical hardening of binaries \citep[e.g.,][]{1994ApJ...431..231V,2007ApJ...658.1047F}.  
It was also realized that even a small primordial 
binary fraction can support the core from deep collapse for more than a Hubble time \citep{2007ApJ...658.1047F}.  
However, comparison between theoretical predictions and observations show that the theoretically predicted 
core radii during the binary-burning phase for these clusters 
are at least an order of magnitude smaller than the observed core radii for the bulk of the GGCs \citep{1994ApJ...431..231V}.  
Additional energy sources were proposed to explain these large core sizes 
\citep[e.g.,][]{2007MNRAS.379L..40M,2008MNRAS.tmp..374M,2008IAUS..246..151C,2007MNRAS.374..857T}, 
however, these scenarios 
need rather special conditions and seem unlikely to be satisfied by most of the GGCs.  Very basic 
questions remain.  Does the bimodal 
distribution for the GGC core sizes separating the core-collapsed and non core-collapsed clusters 
indicate a physical difference between these clusters?  What dynamical stage are the core-collapsed and 
non core-collapsed clusters in today?  Can the large core sizes of the non core-collapsed (majority of the 
Galactic population) be explained without any special conditions simply as a result of $\approx 12\,\gyr$ of evolution 
from realistic initial conditions?

In this study we present the results from a large number of computer simulations (224) with initial
conditions drawn from a multidimensional grid, spanning all relevant parameter ranges
as suggested by observations of young star clusters,
with the goal of reproducing a population of old clusters similar to the GGCs.  We use 
CMC, a \henon-type Monte Carlo code including all physical processes such as single and binary stellar evolution, 
two-body relaxation, strong encounters comprising physical collisions and binary-mediated scattering, and tidal stripping
due to the Galactic tidal field. CMC has been extensively tested and the results from CMC show excellent agreement with those 
from direct $N$-body simulations \citep[e.g.,][]{2007ApJ...658.1047F,2010ApJ...719..915C,2012ApJ...750...31U}. Our goal is 
to find whether starting from realistic initial conditions (including $N$, stellar mass function, central density, compactness 
parameter, binary fraction, and cluster size) typical of the observed young massive 
star clusters \citep[e.g.,][]{2007A&A...469..925S,2009gcgg.book..103S} and without any 
special treatment clusters similar in properties to the observed old GGCs are naturally obtained after 
$\approx 12\,\gyr$ of evolution.  In particular, we focus on understanding the bimodal 
distribution of the observed GGC core radii to identify the dynamical stages of the so called 
core-collapsed and non core-collapsed clusters.

In Section\ \ref{sec:numerics} we briefly explain our code and introduce working definitions for 
key structural parameters of a cluster.  In Section\ \ref{sec:initial_conditions} we describe the multidimensional 
grid of initial parameters explored in this study.  
In Section\ \ref{sec:results} we present our key results.  In Section\ \ref{sec:core-collapsed} we show 
our results identifying the dynamical evolutionary state for the observed core-collapsed GGCs.  Finally we conclude 
in Section\ \ref{sec:conclusion}.

\section{Numerical Method}
\label{sec:numerics}
We use our \henon-type Cluster Monte Carlo code (CMC) to numerically model 
star clusters with single and binary stars including all physical processes relevant in globular clusters 
such as two-body relaxation, single and binary stellar evolution, strong encounters including physical 
collisions and binary mediated strong scattering encounters.  This code has been developed and 
rigorously tested over the past decade \citep{2000ApJ...540..969J,2001ApJ...550..691J,2003ApJ...593..772F,2007ApJ...658.1047F,2010ApJ...719..915C,2012ApJ...750...31U,2012arXiv1206.5878P}.  Using a large grid of initial conditions over the range 
of values typical for observed young massive clusters \citep[e.g.,][]{2007A&A...469..925S,2009gcgg.book..103S} 
we create over $200$ detailed star-by-star models and evolve them
for $12\,\rm{Gyr}$.  A handful (about 6) of these models 
reach a very deep collapse phase.  For these models the CMC time steps become minuscule and the code grinds to a 
halt.  We stop our simulations at that point for these clusters.  In reality, the deep-collapse 
phase is halted via formation of the so-called ``three-body" binaries and the cluster 
enters into the gravo-thermal oscillation phase.  Since in CMC we do not yet include the possibility 
of creating new binaries via three body encounters, we do not address this phase at this stage.  
This however, is not a serious limitation.  All clusters that reach a very deep collapse stage before the integration 
is stopped at $12\,\gyr$, had zero primordial binaries which is not realistic \citep[e.g.,][]{2008AJ....135.2155D}.  
Here, we only include these clusters in our analysis as limiting cases.    
Even a small non-zero primordial binary fraction ($f_b$; lowest $f_b = 5\%$ is used in our simulations) can 
stop the cluster core from collapsing through the dynamical hardening of binaries preventing very deep core collapse. 
At this stage the 
core size remains more or less constant, which is also commonly referred to as ``binary-burning" stage.  
A more detailed description of the code, 
and the various qualitatively different dynamical stages for a cluster's evolution  is
presented in \citet{2010ApJ...719..915C}.

Since one key goal for this study is to compare the properties of our simulated models at about $12\,\gyr$ with 
the properties of observed GGCs, we have to make sure the same parameter definitions are used.  
In particular there are different definitions for the core radius and the cluster size
commonly used by theoreticians and observers. The three dimensional core radius $r_c$ 
widely used in $N$-body simulations is a density-weighted measure related to the virial radius in 
the core \citep{1985ApJ...298...80C}. In contrast, the core radius for observed clusters, $\rcobs$, is often defined 
as the distance from the center where the projected surface brightness profile drops to half its central value.
Alternatively, when observations with sufficiently high resolution are available, the core radius can also be
based on star-counts representing the radius where the projected surface number density is half the corresponding central value.
If the core radius is calculated using the stellar number density profile, 
we will call it $\rcn$, while the core radius based on the brightness, or luminosity, density profile 
is denoted by $\rcl$.  In real clusters the 
density profiles can have large scatter, especially close to the centre due to Poisson noise and the presence 
of bright giants that dominate the light but are only few in number. Hence, it is common practice 
to exclude the light from the brightest giants when calculating the luminosity density profile for real clusters 
to reduce noise \citep[e.g.,][]{2006AJ....132..447N}. For the purpose of this study we define 
$\rclcut$ as the core radius obtained from a luminosity density profile excluding giants with $L_\star > 20\,\lsun$.  
Since the estimated value of $\rcobs$ is strongly dependent on the estimate of the peak luminosity density at the 
centre of a cluster, values of $\rcl$ and $\rclcut$ can differ significantly.     
For real clusters even when star counts are available, only stars above a certain brightness 
can be counted due to completeness issues. Hence, we define $\rcncut$ calculated using a number density profile 
including only stars that are on the main-sequence or on the giant branch with masses $> 0.2\,\msun$, representing the
currently achievable completeness limit \citep[e.g.,][]{2012MNRAS.422.1592L}.
We find that the $\rcncut$ values and the $\rcn$ values are not much different for our simulated clusters. 

The three dimensional half-mass radius $r_h$ is defined as the radius that includes half of the total 
mass of the cluster. However, for real clusters this quantity is not directly accessible, rather, the two dimensional, 
projected distance enclosing half of the total light of the cluster is calculated and is defined as the half-light 
radius $r_{hl}$.  We further define $\rhlcut$ for the half-light radius for our simulated models 
by calculating the same quantity but excluding bright giants with $L_\star > 20\,\lsun$.  We find that the 
$\rhl$ and $\rhlcut$ values show only minor differences.   

\section{Initial Conditions}
\label{sec:initial_conditions}
Our choice of initial conditions is based on observations of super star clusters.
All our simulated clusters have initial virial radii between $r_v = 3$ -- $4\,\pc$ independent of other cluster parameters. 
This range corresponds to initial three-dimensional half mass radii, $r_h$, ranging from $2$ -- $3\,\pc$.
These values are in agreement with observations of young, massive clusters that indicate that the effective cluster sizes are
rather insensitive to the cluster 
mass, as well as metallicity \citep[e.g.][]{2001AJ....122.1888A,2007A&A...469..925S,2009gcgg.book..103S} 
and have a median value of $\approx 3\,\rm{pc}$.  In addition, observations of 
old massive LMC clusters, old GCs in NGC\ 5128, old clusters in M51, 
as well as the GGCs indicate that the effective cluster radii show only a weak correlation with 
the distance from the galactic center \citep{1962PASP...74..248H,1984ApJ...287..185H,1984ApJ...276..491H,1987ApJ...323L..41M,1991ApJ...375..594V,2007A&A...469..925S,2008AJ....135.1567H}.             

To restrict the huge parameter space to a certain extent we place all our simulated 
clusters in a circular orbit at a moderate Galactocentric distance of $r_{G}=8.5\,\rm{kpc}$.  
We avoid modeling GCs very close to the Galactic center, where 
the Galactic field is so strong that the tidal stellar loss dominates the cluster's evolution.  
Due to the assumption of spherical symmetry, Monte Carlo codes cannot directly model the 
tear-drop shaped tidal boundary of a star cluster.  Instead these codes use some prescription 
based method \citep[see a detailed discussion and calibration in][]{2010ApJ...719..915C}.  
If tidal dissolution is the dominant driver 
of the GC's evolution, the approximate method may not be accurate enough.
Choosing a circular orbit for the simulated clusters is a simplification; however, the results 
should still be valid for eccentric orbits with some effective Galactocentric distance 
($>8.5\,\rm{kpc}$) \citep[e.g.,][]{2003MNRAS.340..227B}.  
The Galactic tidal field, and consequently the initial tidal radius, ($r_t$), for the clusters is calculated 
following Equation\ \ref{eq:rt} \citep{2003MNRAS.340..227B},      
\begin{equation}
\label{eq:rt}
r_t = \left( \frac{GM_{\rm{cl}}}{2V_G^2}\right)^{1/3} R_G^{2/3}\, , 
\end{equation}
using a Galactic rotation speed $V_G = 220\,\rm{km}\,\rm{s}^{-1}$.  Here the cluster mass is denoted by $M_{\rm{cl}}$.

For our set of runs we vary the initial number of stars, $N_i$, between $4 -10\times10^5$, 
encompassing the bulk of the GGCs \citep{2005ApJS..161..304M}.
The initial positions and velocities are sampled from a King model distribution function with
dimensionless potential, $W_0$,
in the range $4-7.5$. We vary the initial binary fraction, $f_b$, between $0-0.3$.
The stellar masses of the stars, or primaries in case of a binary, are chosen
from the IMF presented in \citet[][their Equations\ $1$ and $2$]{2001MNRAS.322..231K} 
in the stellar mass range $0.1-100\,\rm{M_\odot}$.  
Secondary binary companion masses
are sampled from a uniform distribution of mass ratios in the range $0.1\,\rm{M_\odot}-m_p$, where 
$m_p$ is the mass of the primary.  
The semi-major axes, $a$, for stellar binaries are chosen from a 
distribution flat in log within physical limits, namely between
$5\times$ the physical contact of the components and the local 
hard-soft boundary.  Although initially all binaries in our models are hard at their respective positions, 
some of these hard binaries can become soft during the evolution of the cluster.  
The cluster contracts under two-body relaxation and 
the velocity dispersion increases making initially hard binaries soft.  Moreover, binaries 
sink to the core due to mass segregation where the velocity dispersion is higher than that at the initial binary positions.  
We include these soft binaries in our simulations until 
they are naturally disrupted via strong encounters in the cluster.  

\section{Basic Structural Parameters of Our Model Clusters}
\label{sec:results}
Here we present the evolution of some structural properties of the simulated clusters and compare 
them with the same properties of the observed GGCs. For each 
of these comparison plots, the evolution of a certain cluster property is shown together with the
distribution of the corresponding observationally derived values for the GGC population.
Since all our simulated models are at a Galactocentric 
distance of $8.5\,\kpc$, we also show the observed distributions for
a subset of GGCs satisfying $7 \leq R_{G} \leq 10\,\kpc$.  
Our goal here is to simply test whether, starting from 
observationally motivated initial conditions typical for massive, young clusters, the final properties of 
the cluster ensemble naturally attain ranges of values as observed in the GGCs.
We do not intend to reproduce the present day distribution for these 
properties since this would require to introduce the distribution of cluster initial conditions as another parameter, and,
consequently, significantly more simulations which is beyond the scope of this study.  

\begin{figure}
\begin{center}
\includegraphics[width=0.9\textwidth]{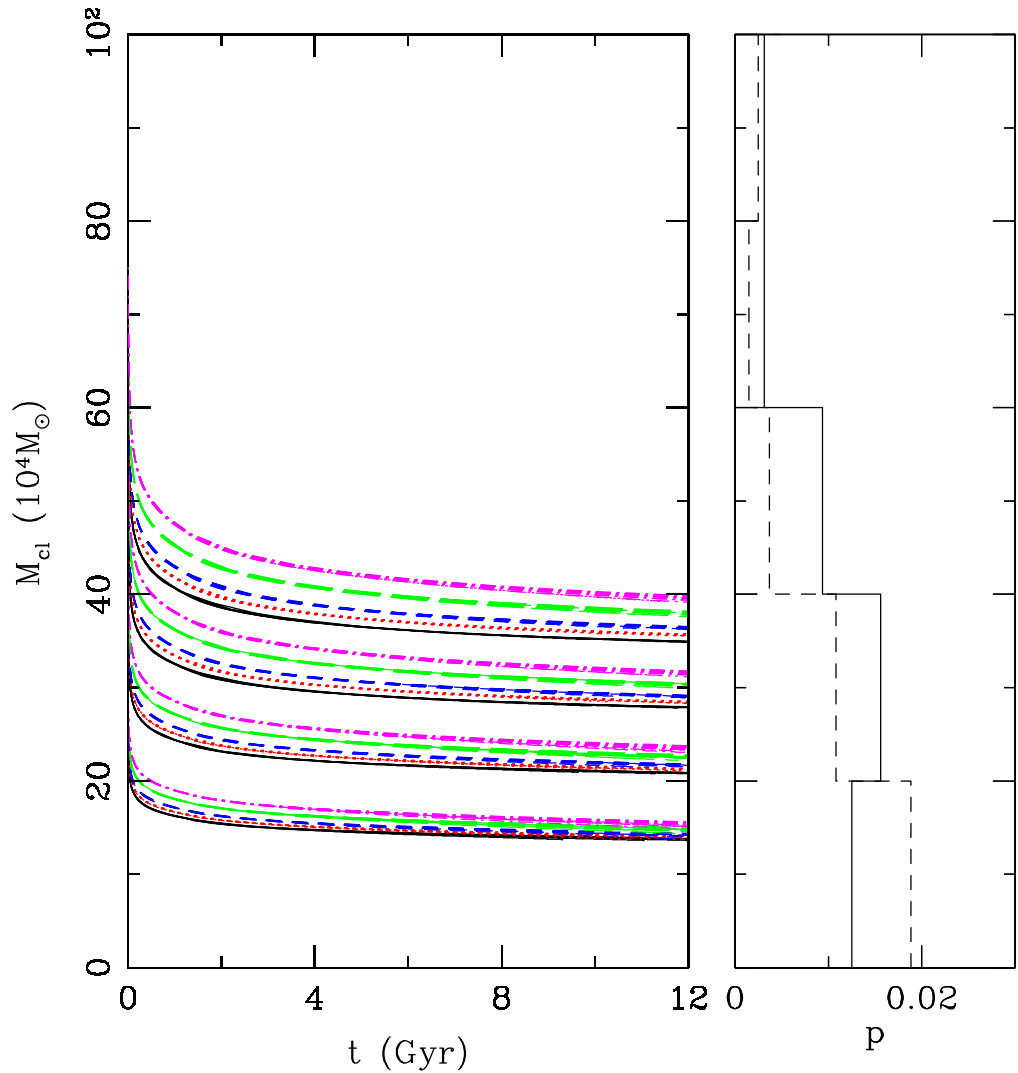}
\caption[$t$ vs $M_\rm{cl}$]{Evolution of the star cluster mass for all simulated models.  
The solid (black), dotted (red), short-dashed (blue), long-dashed (green), and dot-dashed 
(magenta) lines denote models with primordial $f_b = 0$, $0.05$, $0.1$, $0.2$, and $0.3$, 
respectively.  The histograms show the mass distributions of the observed GGCs.  The solid histogram 
is for GGCs with Galactocentric distances between $7$ and $10\,\kpc$.  The dashed 
histogram is for all observed GGCs.  The masses of the observed GGCs are derived from 
their integrated V-band magnitudes \citep{1996AJ....112.1487H} using Equation\ \ref{eq:M/L}.   }
\label{plot:t_mass}
\end{center}
\end{figure} 

Figure\ \ref{plot:t_mass} shows the evolution of the total mass of our simulated clusters. 
The initial sharp decrease in the cluster mass ($M_{\rm{cl}}$) is because of the high mass loss rate 
via winds and compact object formation of the massive stars in the GC.  Later on, $M_{\rm{cl}}$ 
decreases at a slower, nearly constant rate caused by a steady stellar 
mass loss through the tidal boundary of the GC 
\citep{2003MNRAS.340..227B,2010MNRAS.tmp..844D}.  The histograms on the right show the
distribution of observed GGC masses.  The observed GGC masses 
are estimated from the absolute visual magnitudes ($M_v$) given in 
\citet{1996AJ....112.1487H} 
using Equation\ \ref{eq:M/L} assuming a uniform mass to light ratio $\moverl = 2\,\moverlsun$ for all clusters.  
\begin{equation}
M_{\rm{cl}} = 10^{ - (M_v - 4.75)/2.5 + 0.30103}. 
\label{eq:M/L}
\end{equation}
This is an approximation.  The $\moverl$ can vary from cluster to cluster and can also depend 
on the evolutionary stage of the cluster in question \citep{2009A&A...502..817A}. However, since here 
we are only interested in reproducing the cluster mass ranges for most of the GGCs 
and do not try to model a particular cluster, this estimate should be appropriate.
Figure\ \ref{plot:t_mass} shows that our model clusters have final masses typical for the bulk of the observed GGCs.  
Since these clusters are modeled with the total $N$ and masses typical of the observed GGCs 
the model properties can be directly compared with the overall GGC population without any need 
for scaling.

\begin{figure}
\begin{center}
\includegraphics[width=0.9\textwidth]{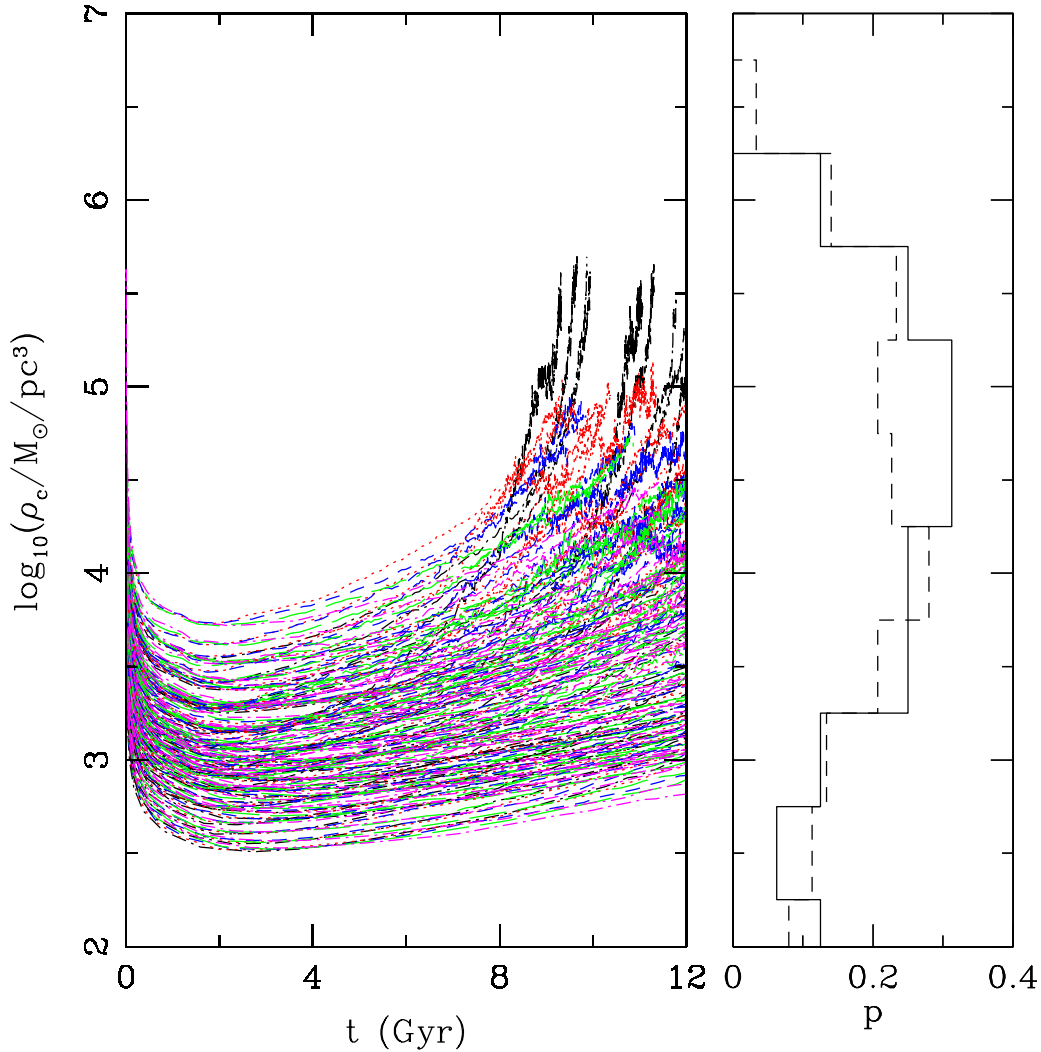}
\caption[$t$ vs $\rho_c$]{Same as Figure\ \ref{plot:t_mass}, but for the central density $\rho_c$.  
The $\rho_c$ values for the observed GGCs are also shown in histograms.  
The solid and the dashed histograms are from GGC populations selected as in 
Figure\ \ref{plot:t_mass}.  
}
\label{plot:t_rhoc}
\end{center}
\end{figure} 
Figure\ \ref{plot:t_rhoc} shows the evolution of the central density in our simulated models.  
As a dynamically important GC property, the central density ($\rho_c$) determines the interaction 
cross-sections for strong scattering inside the core, for example, for BB and BS interactions, 
and stellar collisions. These strong interactions in turn modify the properties of the core, 
through, e.g., binary burning. The $\rho_c$ values sharply 
decrease during the first $\sim 1\,\gyr$ of evolution as the high-mass stars lose mass via 
stellar winds, and compact object formation.  Followed by the sharp decrease 
$\rho_c$ increases almost linearly over time during the gravo-thermal core contraction stage.  
The histograms show the central densities of the observed GGCs.  Here we convert the bolometric 
luminosity densities presented in \citet{1996AJ....112.1487H} and assume $\moverl = 2\,\moverlsun$.  The range 
of central densities for our collection of simulated models is compatible with the range for 
observed GGCs.

\begin{figure}
\begin{center}
\includegraphics[width=0.9\textwidth]{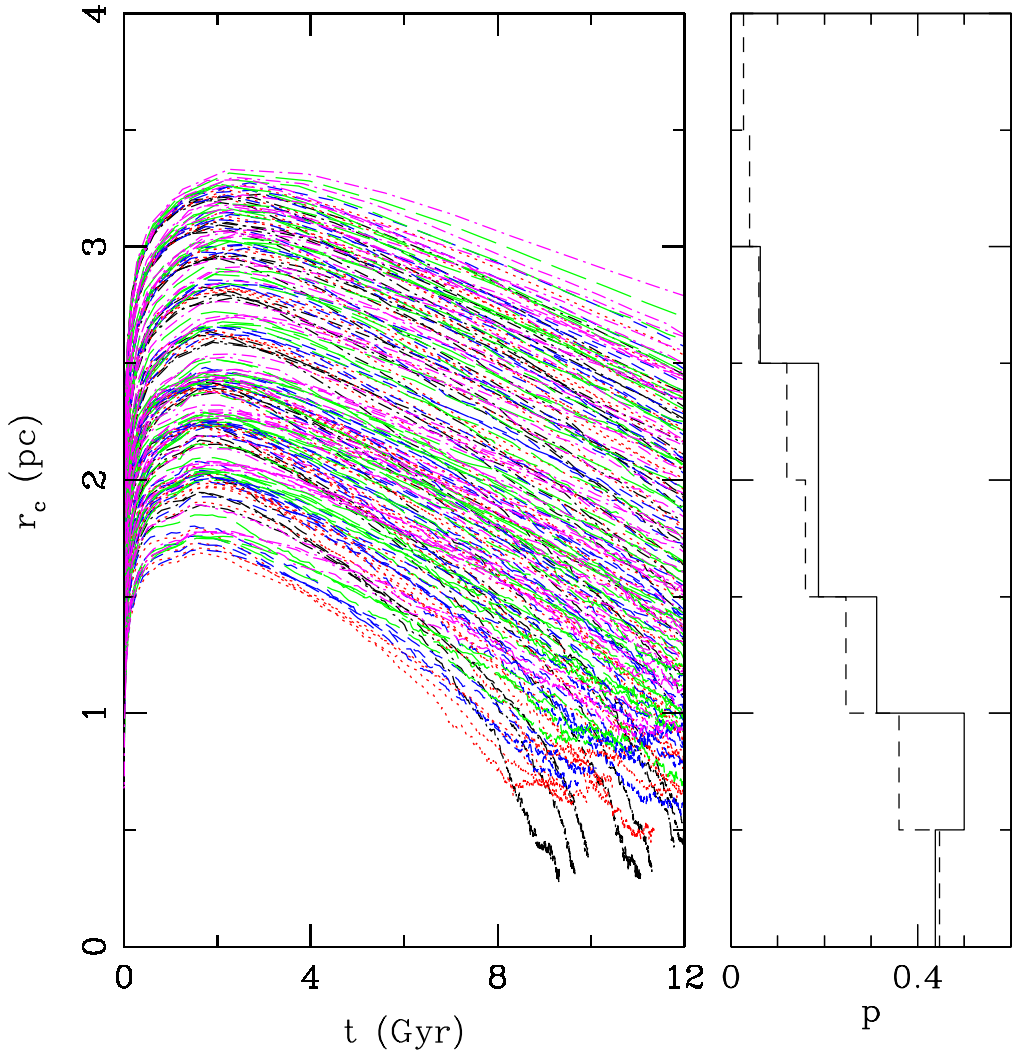}
\caption[$t$ vs $r_c$]{Evolution of the three dimensional core radius $r_c$ for all 
simulated clusters with primordial binary fraction $f_b = 0$--$0.3$.  The line colors 
and styles have the same meaning as in Figure\ \ref{plot:t_mass}.   A few ($6$) 
clusters with no primordial binaries reach the deep collapse phase within the a Hubble 
time.  We stop integrations for those clusters when this phase is reached.  
The histograms show the observed observed core radii for the GGCs.  The solid and dashed 
histograms show GGCs with Galactocentric distances between $7$ -- $10\,\kpc$ and all GGCs.  
}
\label{plot:t_rc}
\end{center}
\end{figure} 
Figure\ \ref{plot:t_rc} shows the evolution of the density weighted three dimensional core radius $r_c$.
The $r_c$ for all clusters sharply increases initially for up to about $1\,\gyr$, because of mass loss due to stellar evolution.   
This mass loss happens mainly at the deepest part of the cluster
potential since the high-mass stars reside near the cluster center due to mass segregation and are affected by mass loss the most.  
The resulting loss 
of gravitational binding energy expands the cluster core.  Once the rate of mass loss goes down, the core 
contracts via diffusive energy transport from the core to the outside through two-body relaxation.  The gravo-thermal contraction 
ceases when this energy flow is balanced by the production of energy via dynamical hardening of binaries, the binary-burning phase.
In our collection of models about 20 clusters show clear binary-burning 
end stages exhibited by a near constant core radius at late stages ($t>8\,\gyr$).  The bulk of our models 
are still contracting at $t_{cl} = 12\,\gyr$.  A few of our models with $f_b = 0$ do not show the binary-burning 
stage and enter deep-collapse after gravo-thermal contraction.  This is because we do not account 
for dynamical creation of binaries.  However, as mentioned before, $f_b = 0$ is a limiting case and not representative for observed GCs. Even with $f_b = 5\%$, we find that the clusters that complete the slow contraction 
phase do not suffer deep collapse, rather reach a steady binary-burning energy equilibrium phase.        

\begin{figure}
\begin{center}
\includegraphics[width=0.9\textwidth]{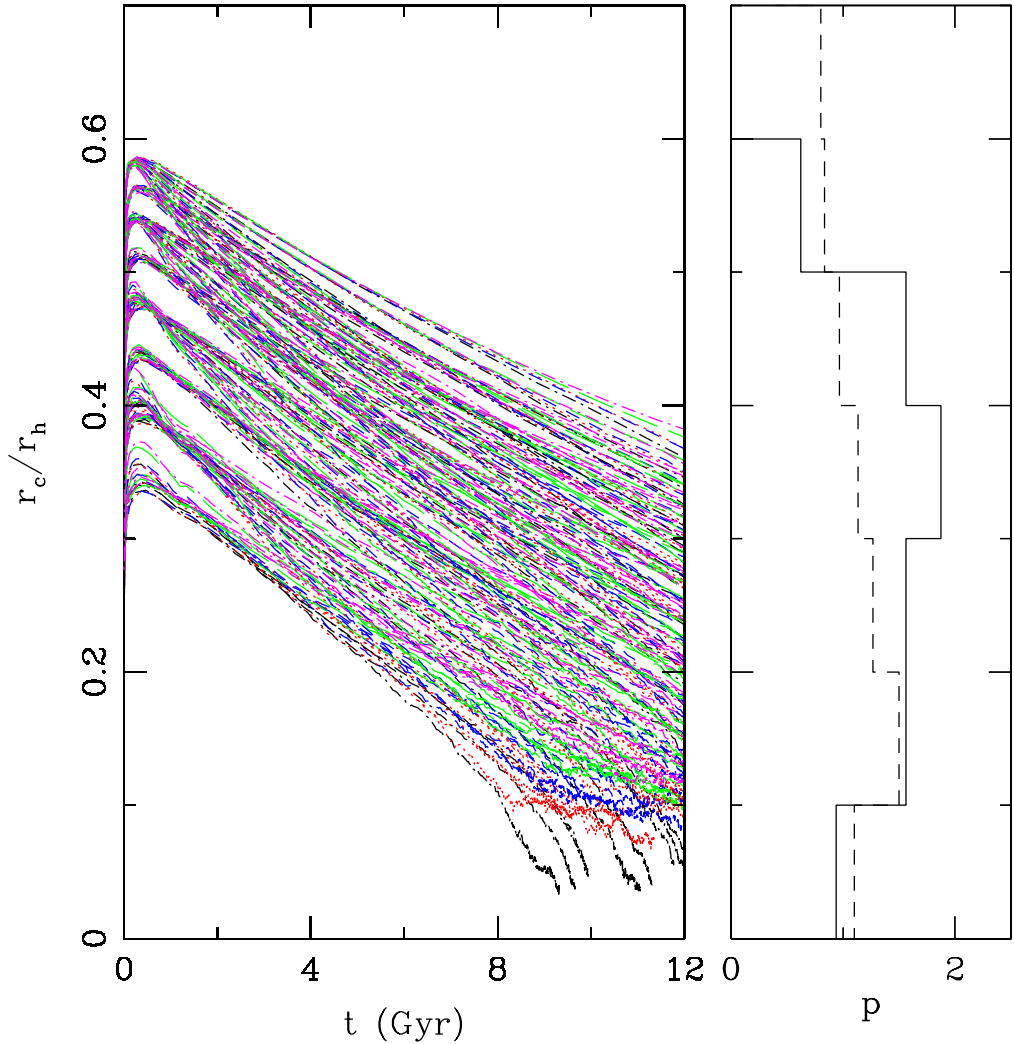}
\caption[$t$ vs $N$-body $r_c/r_h$]{Same as Figure\ \ref{plot:t_rc}, but for the ratio of the 
three dimensional core radius to the three dimensional half mass radius ($r_c/r_h$).  
The $r_c/r_h$ values for the observed GGCs are also shown in histograms.  
The solid and the dashed histograms are from GGC populations selected as in 
Figure\ \ref{plot:t_rc}.  
}
\label{plot:t_rcoverrh}
\end{center}
\end{figure} 
Another reliably measured and frequently used parameter reflecting the dynamical state of the evolution of 
GCs is the ratio between the core radius $r_c$ and the half-mass radius $r_h$ \citep[e.g.,][]{2007ApJ...658.1047F}.  
Figure\ \ref{plot:t_rcoverrh} shows the evolution of $r_c/r_h$ for all our clusters.  
The range of final $r_c/r_h$ values of the simulated clusters agree well with  
the observed ones in the GGC population, producing values at $12\,\rm{Gyr}$ close to 
the peak of the observed distribution.  

\begin{figure}
\begin{center}
\includegraphics[width=0.9\textwidth]{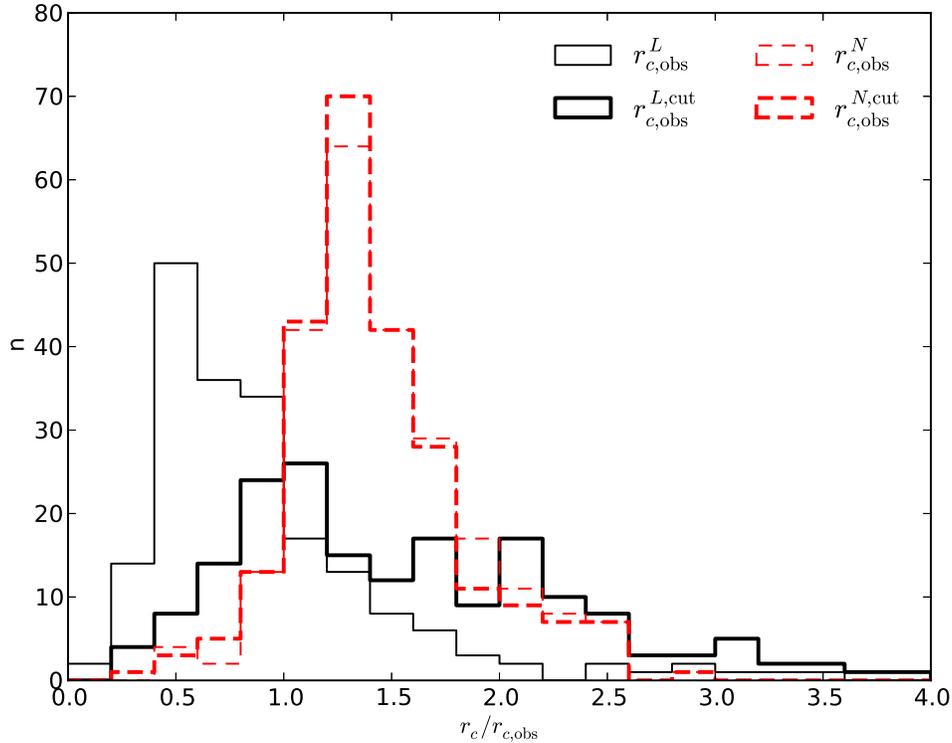}
\caption[$r_c$ and $r_{c, \rm{obs}}$]{Histogram of the ratio of the three dimensional code 
defined core radius $r_c$ to the observed two dimensional projected core radius $r_{c, \rm{obs}}$ 
for all simulated clusters.  The observed core radii $r_{c, \rm{obs}}$ are estimated by using the 
distance from the center of the cluster where the density is half of the maximum density 
at the center.  The different histograms show the distribution of values based on different methods 
to calculate the density distribution.  The thin lines are for observed core radii values calculated 
using a density distribution including all stars in the cluster.  Thick lines are for the same but here 
the density distribution is calculated using a subset of stars satisfying $0.01 \lsun< L_\star < 20\,\lsun$.  
These values ensure that only main-sequence stars $> 0.2\,\msun$ are included and very bright 
giants are excluded.  The solid (black) and dashed (red) histograms are for observed core radii 
values calculated from the stellar luminosity density, and number density distributions, respectively. 
}
\label{plot:rc_rcobs_hist}
\end{center}
\end{figure} 
We should remind the readers, however, that these $r_c$ and $r_h$ values are not 
exactly the quantities observed directly.  A more careful comparison between these structural parameters in 
our models and the observed values in the GGCs requires ``observing" the models and defining the 
model quantities such as $r_c$ and $r_h$ as the observers define them for a real cluster (Section\ \ref{sec:numerics}).  
For example, at $12\,\gyr$ our model {\tt run12} has the $N$-body defined $r_c = 2.6\,\pc$.  If the same model is observed 
and the core radius is calculated using the surface number density distribution using all main-sequence stars satisfying 
$M_\star \geq 0.2\,\msun$ and low-luminosity ($L_\star < 20\,\lsun$) 
giant stars, the core radius $\rcncut = 2.0$ (Table\ \ref{tab:list}).  Figure\ \ref{plot:rc_rcobs_hist} 
shows the distribution of values for the ratio $r_{c, \rm{obs}} / r_c$ using $r_{c, \rm{obs}}$ values obtained in 
4 different ways.  The values for the different definitions of $r_{c, \rm{obs}}$ and the resulting ratios follow different distributions.  
Thus, if one wants to estimate what the three dimensional $N$-body defined $r_c$ value 
is from the observed value of $r_{c, \rm{obs}}$ and vice versa, one should be careful about how $r_{c, \rm{obs}}$ for that cluster 
was calculated.  In general, $r_{c, \rm{obs}}$ calculated using the number density profile of the cluster show lower 
levels of errors.  Moreover, the distributions do not change significantly based on which stars were included in 
the sample to calculate the surface number density profile. The code defined $r_c$ is typically between $1$ -- $2$ 
times the $r^N_{c, \rm{obs}}$ for all cluster models in our collection.  
The luminosity density profiles are subject to a lot more noise compared to the number density profiles.  As a result 
the $r_c/r^L_{c, \rm{obs}}$ calculated using the luminosity density profiles show a larger spread.  In addition, which 
stars were included in calculating the luminosity density profile matters in this calculation significantly.  Typically, 
the code-defined $r_c$ values are between $0.25$ -- $1.5$ times and $0.5$ -- $2.5$ times the $r_{c, \rm{obs}}^{L}$ and 
$\rclcut$, respectively.   

\begin{figure}
\begin{center}
\includegraphics[width=0.9\textwidth]{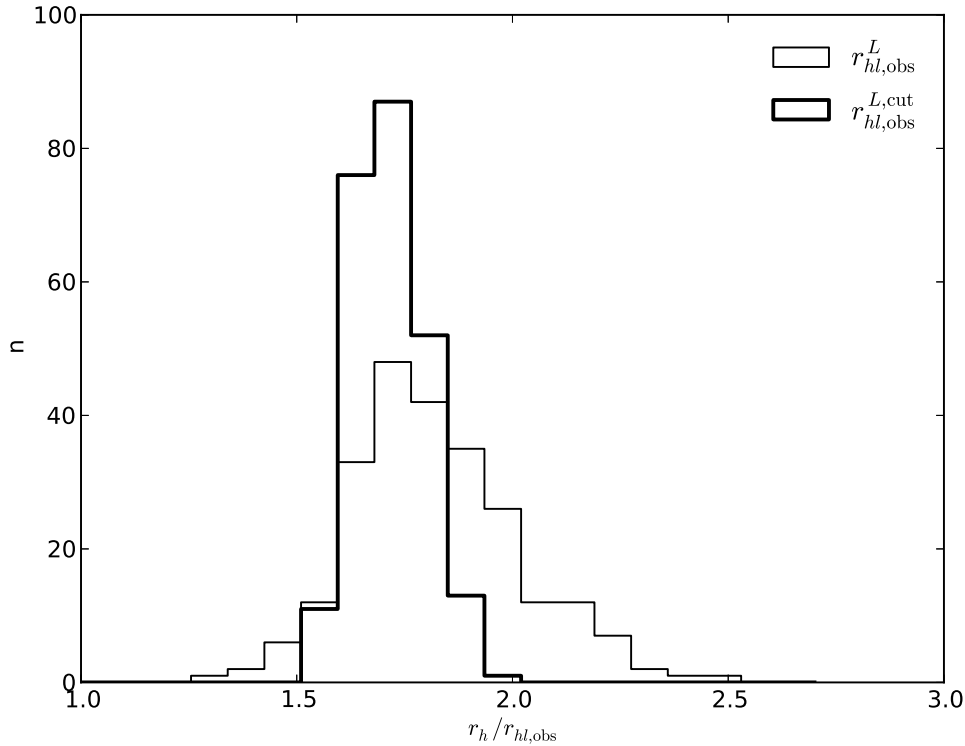}
\caption{Histogram for the ratio of the three dimensional code defined half-mass radius $r_h$ 
to the two dimensional observed half light radius $r_{hl, \rm{obs}}$.  The thin and thick lines 
have the same meaning as in Figure\ \ref{plot:rc_rcobs_hist}.  
}
\label{plot:rh_rhlobs_hist}
\end{center}
\end{figure} 
Similarly, only the half-light radius projected on the sky ($r_{hl}$) is observed in reality and not the three dimensional half 
mass radius ($r_h$).  For example, for the simulated cluster {\tt run12} the sky projected half light radius including all stars is 
$r_{hl, \rm{obs}} = 4.1\,\rm{pc}$.  If a luminosity cut-off is used
for the same cluster the half light radius becomes $\rhlcut = 4.4\,\rm{pc}$.  The theoretically calculated three dimensional 
half-mass radius for the same cluster at the same age is $r_h = 7.0\,\rm{pc}$ (Table\ \ref{tab:list}).  
This difference is not simply due to the projection effect.  This difference depends intricately on the positions of the 
bright giant stars that dominate the light but are few in number.  Figure\ \ref{plot:rh_rhlobs_hist} shows the 
distribution of the ratio $r_h/r_{hl}$ calculated using two separate samples of stars, i.e., with and without applying a luminosity cut-off. Typically, the three 
dimensional half mass radius $r_h$ is between $1.5$ -- $2$ times the observed $\rhlcut$ 
if the luminosity cut described in Section\ \ref{sec:numerics} is used.  Note that this range includes the expected geometric factor of 
$\sqrt{3/2}$ due to projection effect.  The spread in values is moderately larger for  $r_h/r_{hl, \rm{obs}}$.  
This is due to the increased statistical fluctuations created by the bright 
and low number of giants with $L_\star > 20\,\lsun$.       

We are not the first to point out this discrepancy in definitions. For instance, 
\citet{2007MNRAS.379...93H} and \citet{2010ApJ...708.1598T} already found that
the $N$-body definition of $r_c/r_h$ can differ from an 
observed $r_c/r_h$ by a factor of a few, and our results confirm that. 
However, Figures\ \ref{plot:rc_rcobs_hist} and 
\ref{plot:rh_rhlobs_hist} are expected to be useful for observers and theorists modeling GCs since 
the three dimensional $r_c$ or $r_h$ can be be easily estimated using the presented distributions if the 
observed values for these are known.    

\begin{figure}
\begin{center}
\includegraphics[width=0.9\textwidth]{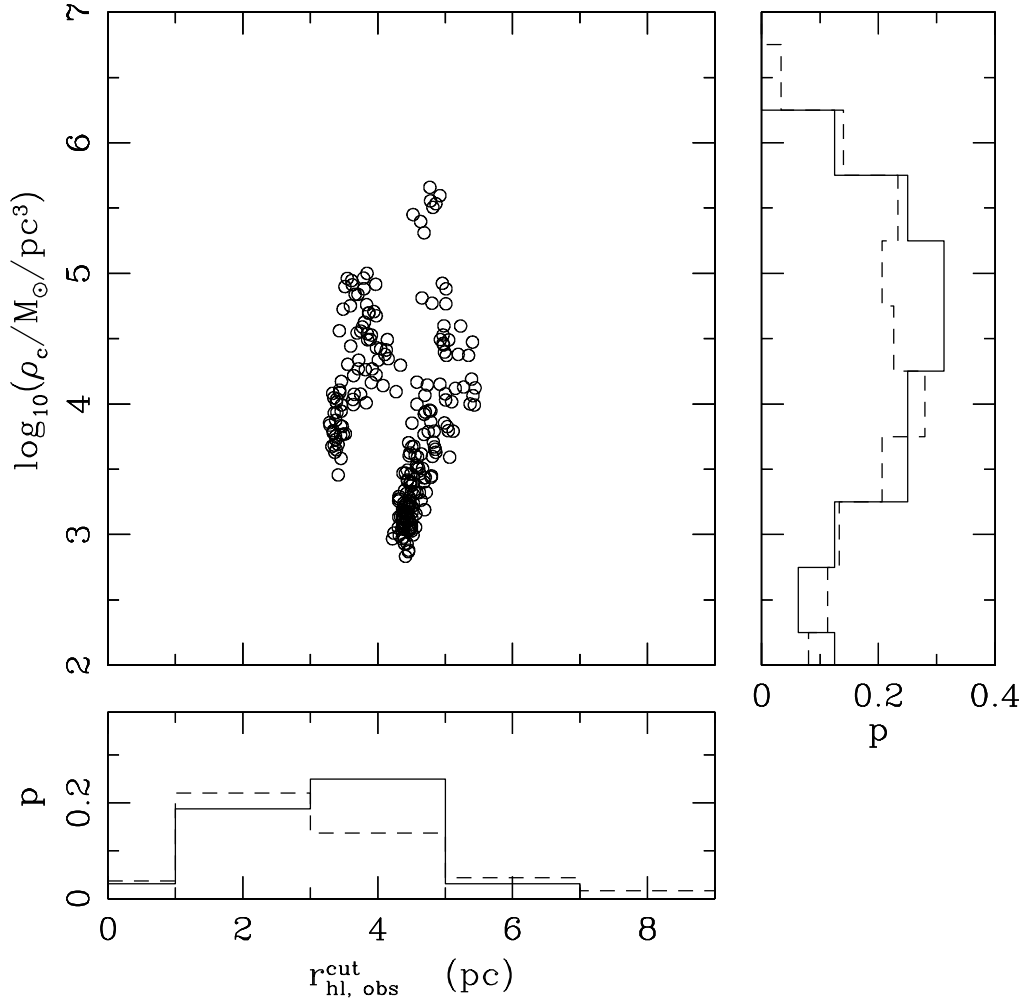}
\caption{The observed half-light radius $r_{hl, \rm{obs}}^{\rm{cut}}$ vs the central density ($\rho_c$) 
for all simulated clusters.  Histograms on either side 
show the $r_{hl, \rm{obs}}$ and $\rho_c$ distributions for the observed GGCs.  The $r_{hl, \rm{obs}}$ and 
$\rho_c$ distributions in our models agree well with the bulk of the observed GGC values.    
The solid and the dashed histograms have the same meaning as in Figure\ \ref{plot:t_rc}.  
}
\label{plot:rh_rhoc}
\end{center}
\end{figure} 
We now compare the values of the structural parameters for our simulated models 
calculated in a similar way as they are calculated for observed GGCs. Figure\ \ref{plot:rh_rhoc} 
shows the $\rho_c$ as a function of $r^{\rm{cut}}_{hl, \rm{obs}}$.  The observed values for all GGCs 
and GGCs in the Solar neighborhood are also shown in the histograms along the respective axes.  
The two clusters of points show $r^{\rm{cut}}_{hl, \rm{obs}}$ vs $\rho_c$ values for clusters with 
initial $r_v = 4$ and $3\,\pc$, respectively.  Our collection of models shows very similar half-light radii 
and central densities compared to those properties for the bulk of the GGCs.  Note that the final $\rhlcut$ 
values are strongly dependent on the initial $r_v$ for a given Galactocentric distance ($r_{G}$) as evidenced in 
the clustering of the points in Figure\ \ref{plot:rh_rhoc}.  Interestingly, for GCs at $r_{G}\approx8.5\,\kpc$ 
the half-light radii remain close to the initial $r_v$ according to our models. Thus, the present observed 
half-light radii can be used as an effective indicator for what the initial $r_v$ was for these clusters.  
The other parts of the observed histograms for $\rhlcut$ can be easily populated if more simulations are performed 
using smaller initial $r_v$.  In addition, using smaller $r_{G}$ will also lead to smaller final $\rhlcut$ values 
for these clusters as they cannot expand further once they fill their tidal radii.

\begin{figure}
\begin{center}
\includegraphics[width=0.9\textwidth]{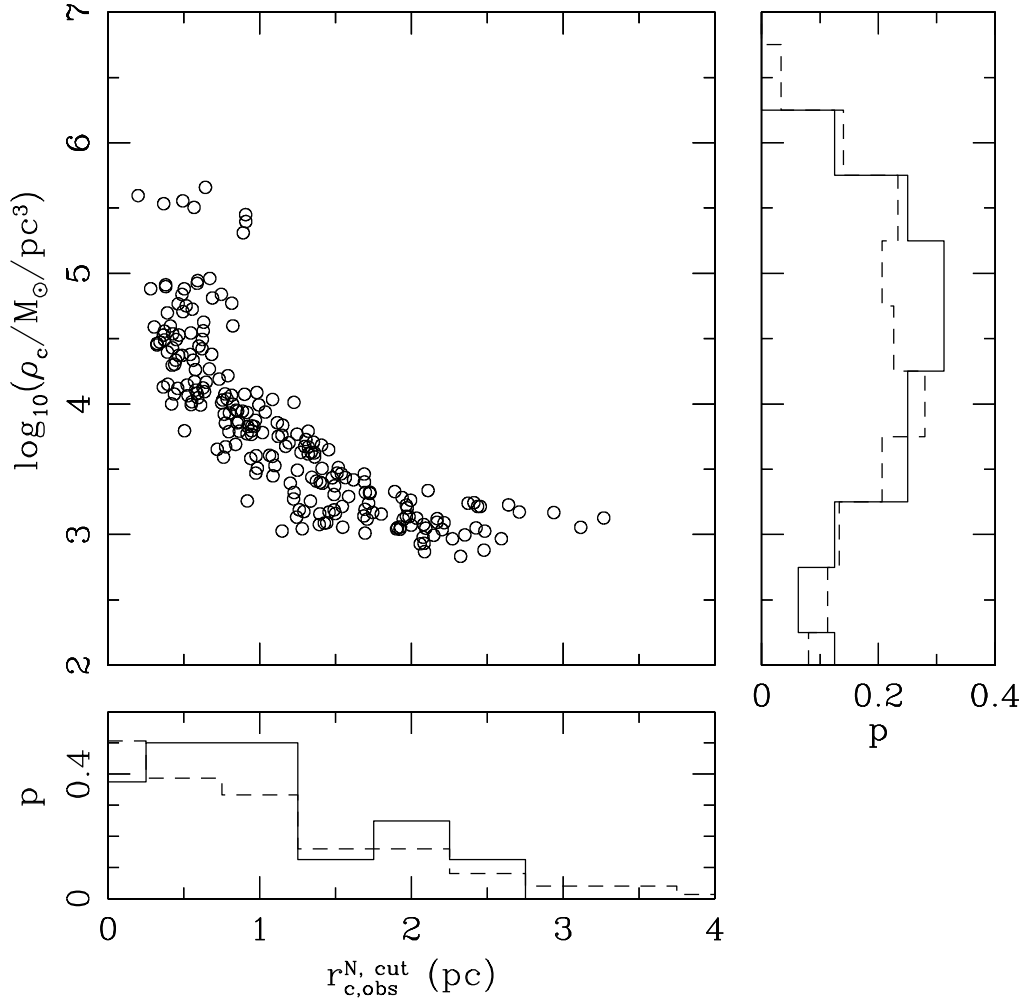}
\caption{$\rcncut$ vs $\rho_c$.  The solid and dashed histograms have the same 
meaning as in Figure\ \ref{plot:t_rc}.  }
\label{plot:rc_rhoc}
\end{center}
\end{figure} 
Figure\ \ref{plot:rc_rhoc} compares the final values of $\rcncut$ and $\rho_c$ for all our simulated models 
with the observed core radius distribution for the GGCs.  As expected, the smaller the $\rcncut$, the higher 
the $\rho_c$.  Our simulated models populate a very similar range in $\rcncut$ values as is observed in the 
bulk of the GGCs.  This is also true when other definitions of $\rcobs$ are used for the comparison.     

\begin{figure}
\begin{center}
\includegraphics[width=0.9\textwidth]{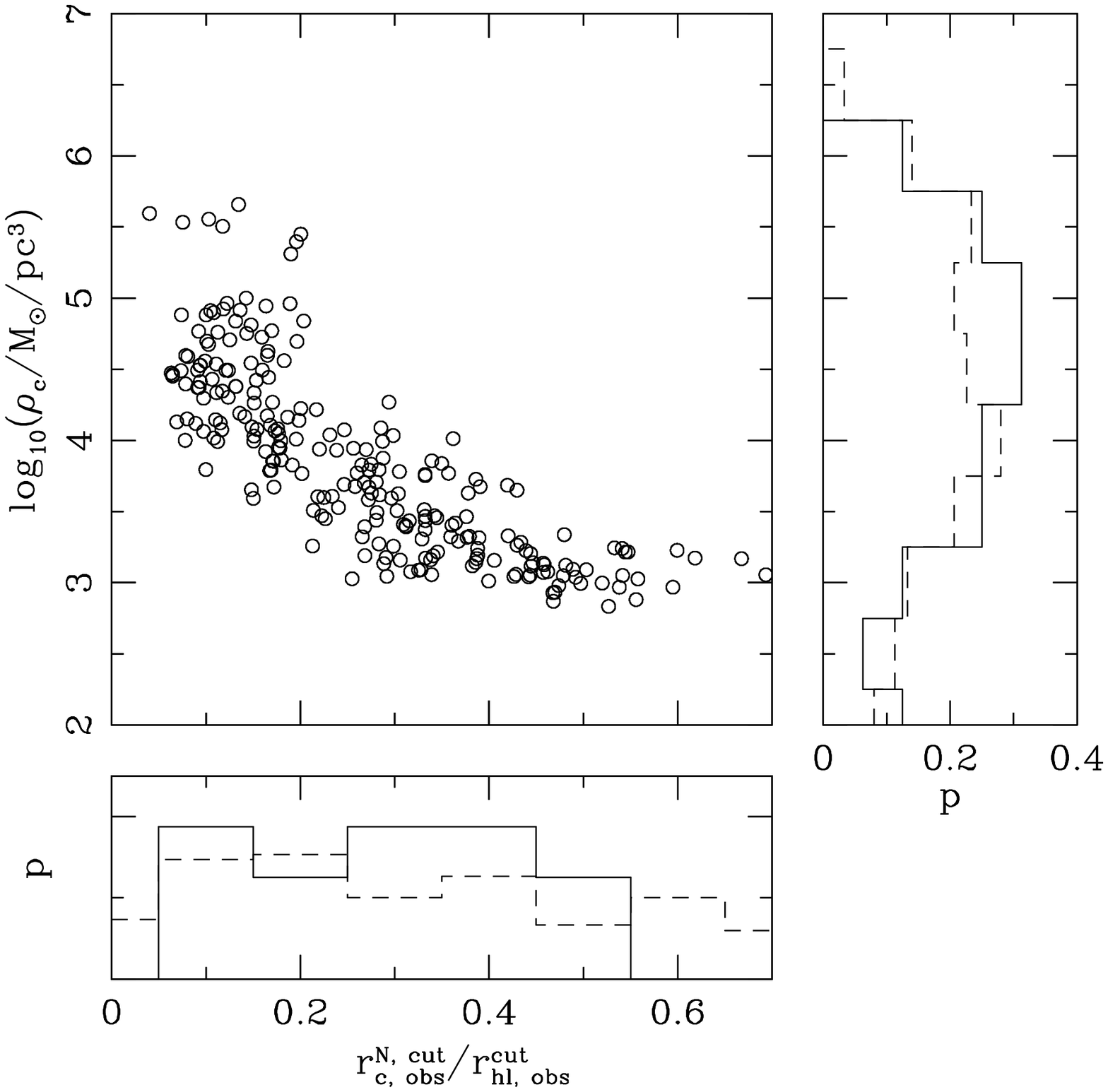}
\caption{Ratio of the observed core radius to the observed half light radius 
$\rcncut / \rhlcut$ vs the central density $\rho_c$.  
The histograms have the same meaning as in Figure\ \ref{plot:rh_rhoc}.     }
\label{plot:rcoverrh_rhoc}
\end{center}
\end{figure} 
Figure\ \ref{plot:rcoverrh_rhoc} shows the scatter plot for the  $\rcncut/\rhlcut$ and 
$\rho_c$ for our models together with the corresponding histograms for the observed GGCs.
The values for this dynamically important and dimensionless measure for the compactness of the cluster
show excellent agreement with the values typically found in all observed GGCs.  

\begin{figure}
\begin{center}
\includegraphics[width=0.9\textwidth]{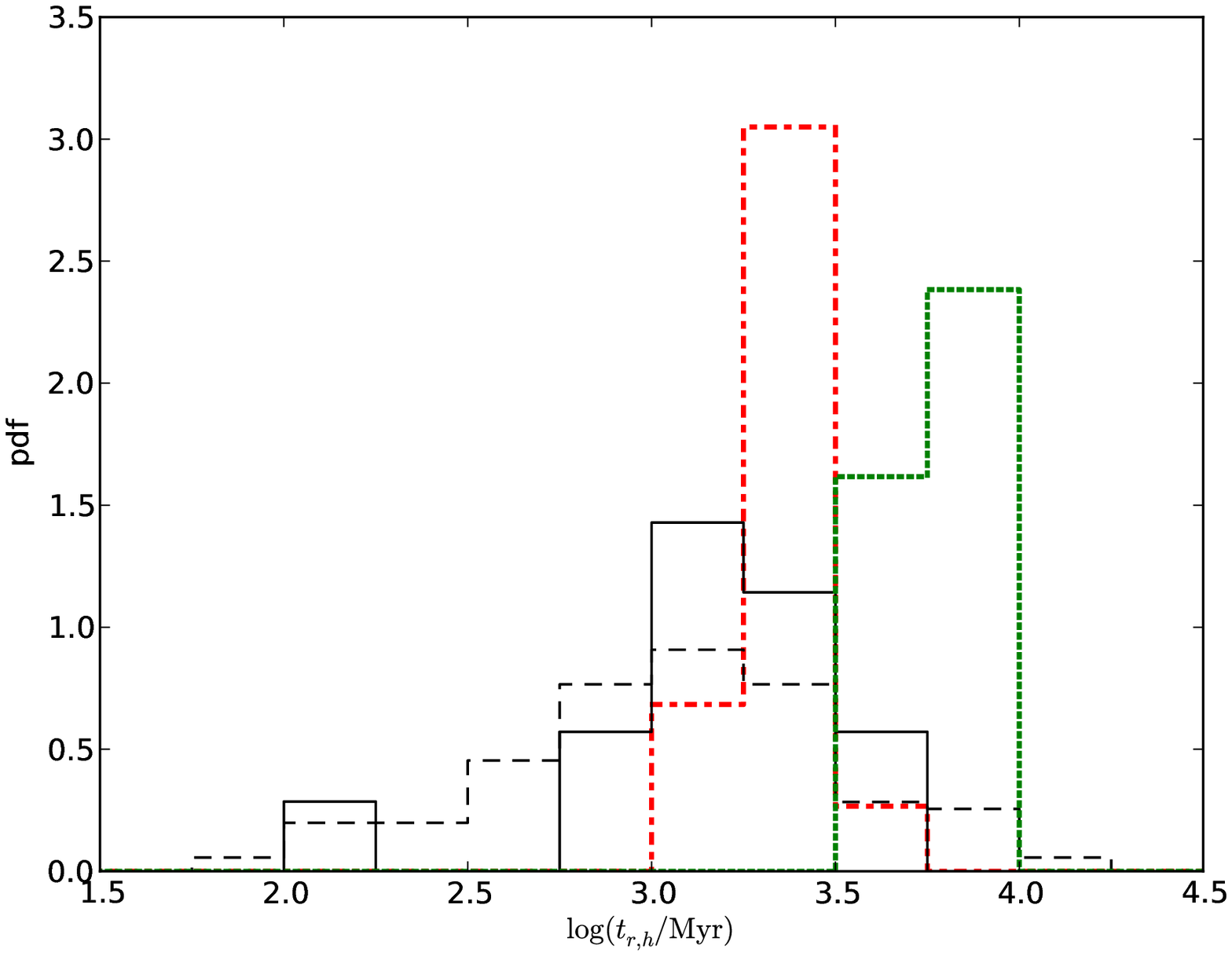}
\caption{Comparison of the half-mass relaxation time $t_{r,h}$ for our models with the observed 
GGCs.  Solid (black) and dashed (black) histograms show the distributions for the $t_{r,h}$ values 
of the GGCs in the Solar neighborhood ($7 \leq r_G \leq 10\,\kpc$), and all GGCs, respectively.  
The dash-dotted (red) histogram is for the $t_{r,h}$ values from our models calculated being consistent 
with the assumptions made in estimating $t_{r,h}$ values in the Harris catalog of GGCs \citep[see text; ][]{1996AJ....112.1487H}.  
The dotted histogram (green) shows the $t_{r,h}$ values from our models if they are calculated using 
the actual values of $M_{\rm{cl}}$, $<M_\star>$, and $r_h$.  The bin at $\log(t_{r,h}/\rm{Myr})$ just above 2 
for the histogram of the Solar neighborhood clusters consists of a single rather unusual sparse cluster E3.  
The model $t_{r,h}$ values show good agreement with the $t_{r,h}$ values of the bulk of the Solar neighborhood 
GGCs.  
}
\label{plot:trh_comp}
\end{center}
\end{figure} 
Figure\ \ref{plot:trh_comp} shows the distributions of relaxation times at $r_h$ ($t_{r,h}$) for our models, 
all GGCs, and a subsample of the GGCs in the solar neighborhood ($7 \leq r_G \leq 10\,\kpc$).  
The relaxation time is a very important dynamical quantity because it is the time scale on which the global cluster
properties evolve.  However, it is difficult to derive this quantity 
accurately for observed clusters since it depends on several dynamical cluster properties
that are not directly observable.
Traditionally, multiple assumptions are made to calculate $t_{r,h}$ for observed GGCs \citep[][see the bibliography in the 
online database]{1996AJ....112.1487H}.  To remain consistent with the derived values
in \citet{1996AJ....112.1487H} we adopt the same assumptions to calculate the $t_{r,h}$ values for 
our models.  At the end of the simulation ($\approx 12\,\gyr$) we calculate the total luminosity ($L_{\rm{cl}}$) 
of the model cluster, and compute its total mass using 
$M_{\rm{cl}}/L_{\rm{cl}}=2\, \msun/\lsun$ (actually can have values between $1.2$ -- $2.2\,\moverlsun$ 
in our models).  The total number of stars is then calculated assuming an average stellar mass 
$<M_\star> = 1/3\,\msun$ (actually can have values between $0.35$ -- $0.39\,\msun$ in our models).  
We use $\rhlcut$ as a proxy for $r_h$. We estimate the $t_{r,h}$ values for our models 
using Equation\ 11 of \citet{1993ASPC...50..373D} with the corrected coefficient as mentioned 
in \citet{1996AJ....112.1487H}. The $t_{r,h}$ values for our 
models calculated this way agree well for the GGCs in the Solar neighborhood.  The bin showing 
an unusually low $t_{r,h}$ value just above $\log(t_{r,h}/\rm{Myr}) = 2$ contains a single cluster E3.  
A quick search indicates that E3 is a sparse unusual cluster.  Again we remind 
the readers that here we are only interested in comparing the ranges of $t_{r,h}$ from our models and 
the observed GGCs.
Note that the values for $t_{r,h}$ based on the actual cluster parameters can be a few times 
higher depending on the particular cluster properties. The differences between the actual 
dynamical $t_{r,h}$ values and those derived for the observed clusters come essentially from 
the many assumptions listed above.            

A long standing puzzle has been the apparent discrepancy between the theoretically 
predicted $r_c/r_h$ values and the values observed for the GGCs.  Early numerical simulations 
as well as analytical studies expected that most GGCs are in the binary-burning stage.  However, 
the simulated $r_c/r_h$ values resulting from binary burning
have been found to be about an order of magnitude smaller than that for the bulk of the observed 
population \citep[e.g.,][]{1994ApJ...431..231V,2007ApJ...658.1047F} and it was already known that 
this amount of discrepancy cannot be entirely coming from differences in definitions 
\citep[e.g.,][]{2007MNRAS.379...93H,2010ApJ...708.1598T}.  Consequently, additional energy sources in the core 
to expand $r_c$ have been investigated.  Several studies proposed different 
additional energy generation mechanisms to explain the large observed $r_c/r_h$ values 
\citep[e.g.][]{2007MNRAS.374..857T,2008IAUS..246..151C,2008ApJ...673L..25F,2008MNRAS.tmp..374M}.  However, 
these additional energy sources require rather special conditions.  For example, 
high central densities are required to create a population of high-mass stars via physical collisions that can then suffer 
expedited mass loss via compact object formation \citep{2008IAUS..246..151C}.  On the other hand, 
presence of an intermediate mass black hole also cannot be common for the GGCs \citep{2007MNRAS.374..344T}.   
Our models, in contrast, are generated without any special assumptions using observationally motivated 
initial conditions, and they naturally create a population of model clusters with properties in excellent agreement 
with the bulk of the observed GGC properties.  These results indicate that the progenitors of today's GGCs 
were very similar in properties to the young massive clusters observed, for example, in M~51 
\citep{2007A&A...469..925S,2009gcgg.book..103S}.  Our results also indicate that the majority of today's GGCs 
have not yet reached the binary-burning, energy-equilibrium stage, and are still contracting under energy transport 
via two-body relaxation.      

\section{What is a ``core-collapsed" cluster?}
\label{sec:core-collapsed}
\begin{figure}
\begin{center}
\includegraphics[width=0.9\textwidth]{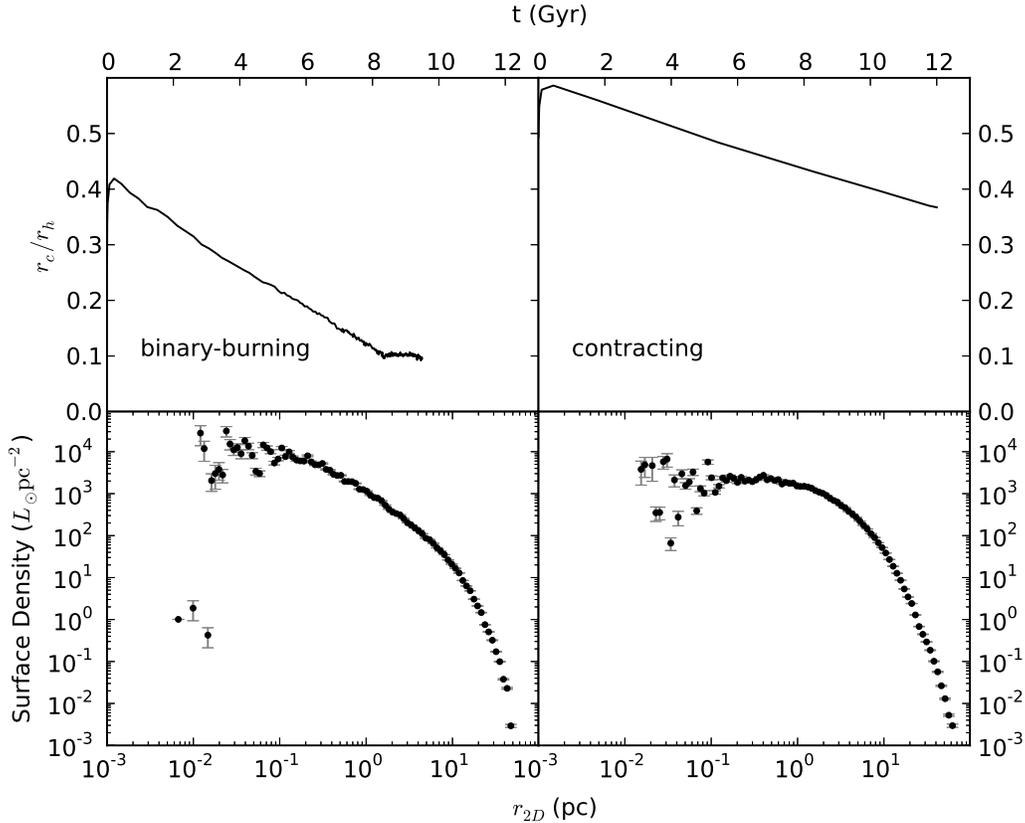}
\caption{Examples for a binary-burning cluster (left) and a core-contracting cluster (right) from our models chosen 
randomly.  The top panels show the evolution of $r_c/r_h$ for each clusters.  The bottom panels show the 
surface luminosity density profiles calculated excluding the bright ($L_\star > 20\,\lsun$) giants to reduce noise.  
The errorbars are estimated $1\sigma$ Poisson errors for each bin.  The binary-burning model shows a clear 
power-law slope until the data is too noisy.  In contrast, the core-contracting model shows a clear King density profile.    
}
\label{plot:example}
\end{center}
\end{figure} 
A lot of early theoretical work was devoted to understand the gravo-thermal collapse and subsequent 
evolution of the core and the cluster as a whole due to hardening of primordial binaries.
However, analytical results as well as numerical simulations showed that the $r_c$ values in the binary-burning phase
are about an order of magnitude too small compared to the bulk of the GGC cores.  

Our simulation results presented in Section\ \ref{sec:results} show that this is simply because the bulk of the 
observed GGCs are {\em not} in binary-burning equilibrium stage.  Rather they are still contracting.  Now we 
focus on understanding the clearly bimodal distribution of the core radii of the GGCs.  Depending on the shape of the 
observed cluster density profiles, all GGCs are divided into two categories, namely core-collapsed and non core-collapsed 
clusters \citep[e.g.,][]{1996AJ....112.1487H,2005ApJS..161..304M}.  
The so called core-collapsed clusters exhibit a power-law increase in the density profile until 
the limit of resolution of the observation.  The non core-collapsed clusters show a very clear flattening of the 
density profile and the profile for these clusters are well fitted with a King profile \citep{1966AJ.....71...64K}.  
Is there a distinct difference dynamically between the two categories of clusters?  

Figure\ \ref{plot:example} shows examples of two representative clusters chosen randomly from our large 
collection of models, one is in a clearly binary-burning stage and the other is still contracting.  The evolution of 
$r_c/r_h$ is shown for the two clusters as well as their respective surface density profiles.  The binary-burning cluster 
can be clearly identified by the near constant value of $r_c/r_h$ starting at about $8\,\gyr$.  The core contracting 
cluster shows a constant rate of contraction until $12\,\gyr$.  The surface density profiles for the two clusters are 
very different.  The surface density profile for the binary-burning cluster looks very much like a so called core-collapsed 
cluster and exhibits a clearly power-law slope to very small radius below which the profile is noisy.  In contrast, the 
surface density profile for the core contracting cluster shows a clear King profile with a distinct flat part near the center below 
a few parsecs.  

\begin{figure}
\begin{center}
\includegraphics[width=0.9\textwidth]{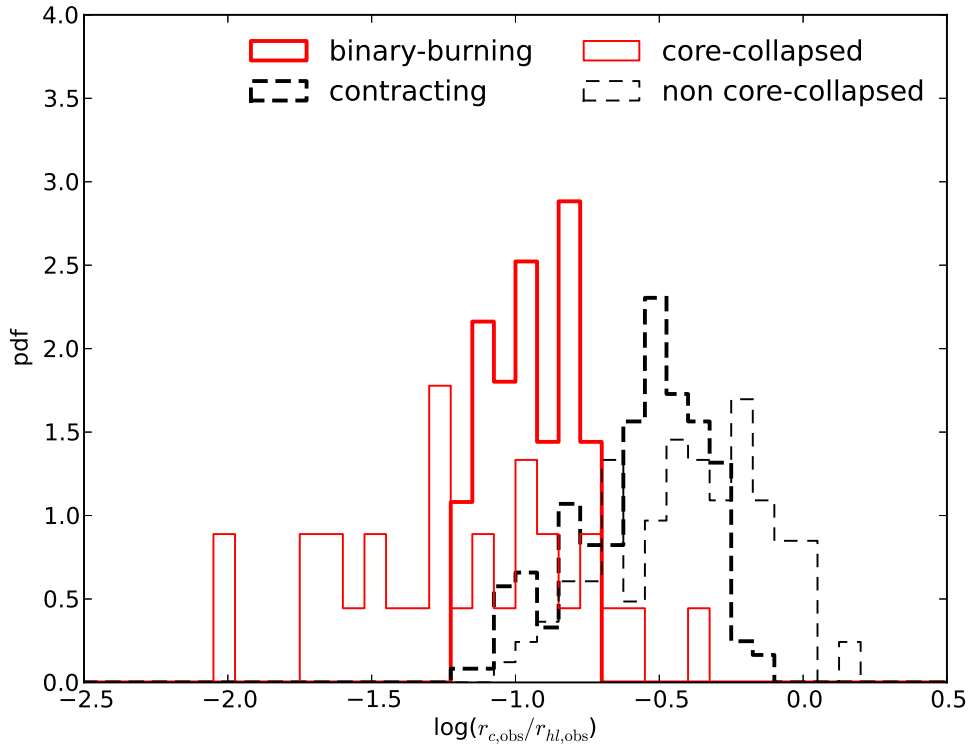}
\caption{Distribution of $r_{c, \rm{obs}}^{N, \rm{cut}}$/$r_{hl, \rm{obs}}^{\rm{cut}}$.  The thick and 
thin lines denote values from our simulated models and observed GGCs, respectively.  The solid 
and dashed histograms show distributions for model clusters in binary-burning energy equilibrium 
and in gravo-thermal contraction stages, respectively.  The solid and dashed histograms for 
the observed GGCs denote core-collapsed clusters and non core-collapsed clusters, respectively.  
The ratio between the core and half-light radius for binary-burning clusters have very similar 
values compared to the values for the core-collapsed population among the GGCs.  On the other 
hand the bulk of the GGCs have values for this ratio similar to the model clusters that are still in the 
gravo-thermal contraction stage at integration stopping time ($\approx 12\,\gyr$).  
}
\label{plot:core-collapsed}
\end{center}
\end{figure} 
We divide the full sample of our simulations in two subgroups: 1) binary-burning: models showing a clear binary-burning 
stage before integration is stopped, and 2) core-contracting: models still contracting due to two-body relaxation until 
integration stopping time.  Figure\ \ref{plot:core-collapsed} shows the distributions for $\rcncut/\rhlcut$ for the two subgroups 
of our theoretical models.  The distributions peak at different values, with the binary-burning clusters having much lower 
core radii.  The distributions for $\rcobs$ for the observed core-collapsed and non core-collapsed GGCs are 
also shown.  Note that the values for the binary-burning clusters from our models and the core-collapsed clusters 
from the observed GGCs show a very similar range of values.  Similarly, $\rcncut/\rhlcut$ for the contracting clusters in our 
models show very similar values compared to those of the observed non core-collapsed GGCs.  In addition, 
most of the core-contracting models with small $\rcncut < 0.1\,\pc$ values are in fact about to start binary-burning, but was 
classified by contracting since the constant $r_c/r_h$ stage of evolution was not clearly seen.  The observed 
GGCs show larger ranges for the core radii values compared for both categories of clusters compared to our models.  
This is simply because we are forced to limit the ranges of the grid of initial conditions to constrain the number of required cluster calculations to a tractable amount.  
We remind the readers that the heights of the histograms between the model clusters and the observed GGCs are not 
compared here, since that depends directly on the distributions of initial cluster parameters, determination of which is beyond the 
scope of this study, and are chosen arbitrarily. Only ranges in values are compared.             

\section{Conclusions}
\label{sec:conclusion}
We have presented a large ($\sim 200$) collection of cluster models created 
with the Northwestern group's  \henon-type Monte Carlo code in star-by-star detail. We start 
our models with initial parameters including the stellar mass spectrum, cluster size, concentration, and primordial binary 
fraction $f_b$ over large ranges (Section\ \ref{sec:initial_conditions}) guided by observed young massive clusters 
\citep[e.g.,][]{2007A&A...469..925S,2009gcgg.book..103S}.  We find that these initial clusters very naturally 
produce a population of model clusters with structural properties including cluster mass, $\rho_c$, $r_c$, $r_h$, 
and $t_{r,h}$ in excellent agreement with the bulk of the 
GGC properties after about $12\,\gyr$ of evolution without any special considerations or fine tuning (e.g., very high 
density to aid collisional expedited stellar mass loss via compact object formation; \citealt{2008IAUS..246..151C}; 
or intermediate mass black holes; \citealt{2007MNRAS.374..857T}).  We pay attention to the various 
different commonly used definitions of the structural parameters $r_c$ and $r_h$ and 
calculate these quantities from our models as an observer would for real clusters.  These parameters  
are then compared and found to agree well with the ranges from observed GGCs.
Using our large collection of models 
we also show the distribution of the ratio of the three dimensional code-defined $r_c$ and $r_h$ 
to the corresponding ``observed" values (Figures\ \ref{plot:rc_rcobs_hist}, \ref{plot:rh_rhlobs_hist}). We expect that these 
distributions of ratios 
for the $r_c$ and $r_h$ values will be valuable for observers and theorists alike to convert the values of these 
parameters from one set of definitions to another.  

From the 
evolution of the code-defined three-dimensional structural parameters of all our models, we find that all qualitatively different 
evolutionary stages are observed, in particular, the initial expansion due to stellar evolution driven mass loss, 
core contraction driven by two-body relaxation, and the binary-burning equilibrium stage (for clusters with $f_b>0$) 
driven by a balance between energy production via dynamical hardening of binaries in the core and outward diffusion of 
energy from the core due to two-body relaxation.  
Our results indicate that the progenitors of today's GGCs 
were very similar in properties to the present day young massive clusters 
\citep[observed, for example, in M~51][]{2007A&A...469..925S,2009gcgg.book..103S}.  Of course, the metallicities 
of these progenitors must have been much lower compared to today's massive young clusters.          

After establishing that our collection of cluster models are representative of the observed GGCs we 
investigate the apparently bimodal distribution of the observed core radii of the GGCs created by 
the core collapsed clusters and non core-collapsed clusters.  In particular, we answer the question 
if the core collapsed clusters are dynamically different from the non core collapsed clusters.    
We find that the surface brightness profile for the binary-burning cluster shows a prominent power-law slope near the 
center (Figure\ \ref{plot:example}).  The core collapsed clusters are observationally defined as the clusters that show this distinct 
feature in their surface brightness profiles \citep[e.g.,][]{2005ApJS..161..304M}.  In contrast, a model cluster 
that is still contracting at $12\,\gyr$ shows a surface brightness profile that has a clear flat central part and is 
fitted well by a King profile \citep{1966AJ.....71...64K}.  We further divide our models 
into two subsets, one containing clusters in the binary-burning stage, and the other containing clusters that are 
still contracting at $12\,\gyr$.  We compare the ratio between the core and half light radii 
of our binary-burning and contracting clusters with those 
for the core-collapsed and non core-collapsed clusters in the observed GGCs, respectively.  
We find that the binary-burning $\rcncut/\rhlcut$ values in our models are in agreement with those of the core-collapsed 
GGCs.  Similarly, the contracting $\rcncut/\rhlcut$ values in our models are in agreement with those of the non core-collapsed 
GGCs (Figure\ \ref{plot:core-collapsed}).  
Thus our results clearly indicate that the so called 
core-collapsed GGCs are in fact at the binary-burning stage 
whereas, the non core-collapsed GGCs are still contracting under two-body relaxation.  This also indicates that 
the majority of the GGCs (since most GGCs are non core-collapsed) are not in energy equilibrium as was expected 
by some earlier theoretical models \citep[e.g.,][]{2007ApJ...658.1047F}.  One key implication for this finding is that 
analytical estimates of interaction rates in a GGC must take into account the fact that the present day observed structural 
parameters including $\rho_c$ has not been constant and is still evolving.  Hence, to calculate a correct estimate one must 
integrate the time dependent cross-sections (based on the changing values of these parameters) over an appropriate length of time 
as has been done in, e.g., \citet{2008ApJ...673L..25F}.   

Acknowledgement: We thank the anonymous referee for his help rectifying some mistakes and useful 
suggestions.  This work was supported by NASA ATP Grant NNX09AO36G at Northwestern University.  
SC acknowledges support from the Theory Postdoctoral Fellowship from UF Department of Astronomy and 
College of Liberal Arts and Sciences.

\begin{landscape}
\begin{center}
\begin{longtable}{cccccccccc|ccccccccccc}
\caption[List of realistic models.]{List of simulations: $W_0$ is the central concentration parameter for a King 
profile \citep{1966AJ.....71...64K}, 
cluster mass $M$ is in $10^5\,\rm{M_\odot}$, 
number of bound cluster objects $N$ is given in $10^5$,   
central stellar mass density $\rho_c$ is in $10^3\,\rm{M_\odot/pc^3}$, 
$r_c$ and $r_h$ are in pc, time of final snapshot $t$ is in Gyr and 
$c$ is the concentration parameter defined as $\log(r_t/r_c)$.  $f_b$ and $f_{b,c}$ 
are the binary fractions in the cluster and that in the core of the cluster, respectively.     
All final values are extracted from the final snapshot of the simulated clusters.
The core radii and half-light radii calculated as they are defined in observed clusters, $\rcncut$ 
and $\rhlcut$ are also listed.  } \label{tab:list} \\   
\hline \hline \\
\multicolumn{1}{c}{\textbf{Name}} & 
\multicolumn{9}{c}{\textbf{Initial}} & 
\multicolumn{11}{c}{\textbf{Final}} \\
\hline\\
\multicolumn{1}{c}{} &
\multicolumn{1}{c}{\textbf{$W_0$}} &
\multicolumn{1}{c}{\textbf{$M$}} &
\multicolumn{1}{c}{\textbf{$N$}} &
\multicolumn{1}{c}{\textbf{$r_c$}} &
\multicolumn{1}{c}{\textbf{$r_h$}} &
\multicolumn{1}{c}{\textbf{$\rho_c$}} &
\multicolumn{1}{c}{\textbf{$f_b$}} &
\multicolumn{1}{c}{\textbf{$f_{b,c}$}} &
\multicolumn{1}{c}{\textbf{$c$}} &
\multicolumn{1}{c}{\textbf{$t$}} &
\multicolumn{1}{c}{\textbf{$M$}} &
\multicolumn{1}{c}{\textbf{$N$}} &
\multicolumn{1}{c}{\textbf{$r_c$}} &
\multicolumn{1}{c}{\textbf{$\rcncut$}} &
\multicolumn{1}{c}{\textbf{$r_h$}} &
\multicolumn{1}{c}{\textbf{$\rhlcut$}} &
\multicolumn{1}{c}{\textbf{$\rho_c$}} &
\multicolumn{1}{c}{\textbf{$f_b$}} &
\multicolumn{1}{c}{\textbf{$f_{b,c}$}} &
\multicolumn{1}{c}{\textbf{$c$}} \\
\hline \\
\endfirsthead
\multicolumn{21}{c}{{\tablename} \thetable{} -- Continued} \\[0.5ex]
\hline \hline \\[-2ex]
\multicolumn{1}{c}{\textbf{Name}} & 
\multicolumn{9}{c}{\textbf{Initial}} & 
\multicolumn{11}{c}{\textbf{Final}} \\
\hline \\
\multicolumn{1}{c}{} &
\multicolumn{1}{c}{\textbf{$W_0$}} &
\multicolumn{1}{c}{\textbf{$M$}} &
\multicolumn{1}{c}{\textbf{$N$}} &
\multicolumn{1}{c}{\textbf{$r_c$}} &
\multicolumn{1}{c}{\textbf{$r_h$}} &
\multicolumn{1}{c}{\textbf{$\rho_c$}} &
\multicolumn{1}{c}{\textbf{$f_b$}} &
\multicolumn{1}{c}{\textbf{$f_{b,c}$}} &
\multicolumn{1}{c}{\textbf{$c$}} &
\multicolumn{1}{c}{\textbf{$t$}} &
\multicolumn{1}{c}{\textbf{$M$}} &
\multicolumn{1}{c}{\textbf{$N$}} &
\multicolumn{1}{c}{\textbf{$r_c$}} &
\multicolumn{1}{c}{\textbf{$\rcncut$}} &
\multicolumn{1}{c}{\textbf{$r_h$}} &
\multicolumn{1}{c}{\textbf{$\rhlcut$}} &
\multicolumn{1}{c}{\textbf{$\rho_c$}} &
\multicolumn{1}{c}{\textbf{$f_b$}} &
\multicolumn{1}{c}{\textbf{$f_{b,c}$}} &
\multicolumn{1}{c}{\textbf{$c$}} \\
\hline \\
\endhead
\multicolumn{21}{l}{{Continued on Next Page\ldots}} \\
\endfoot
\\[-1.8ex] \hline \hline 
\endlastfoot
{\tt run1} & 4 & 2.5 & 4 & 1.6 & 3.3 & 12.2 & 0.00 & 0.00 & 0.84 & 12 & 1.4 & 3 & 2.0 & 1.7 & 7.1 & 4.2 & 1.0 & 0.00 & 0.00 & 1.5 \\ 
{\tt run2} & 4 & 3.8 & 6 & 1.6 & 3.3 & 16.2 & 0.00 & 0.00 & 0.84 & 12 & 2.1 & 5 & 2.3 & 1.9 & 7.0 & 4.3 & 1.1 & 0.00 & 0.00 & 1.5 \\ 
{\tt run3} & 4 & 5.1 & 8 & 1.6 & 3.3 & 23.3 & 0.00 & 0.00 & 0.84 & 12 & 2.8 & 7 & 2.4 & 1.9 & 6.9 & 4.4 & 1.3 & 0.00 & 0.00 & 1.5 \\ 
{\tt run4} & 4 & 6.4 & 10 & 1.6 & 3.3 & 29.0 & 0.00 & 0.00 & 0.84 & 12 & 3.5 & 9 & 2.5 & 2.9 & 6.8 & 4.4 & 1.5 & 0.00 & 0.00 & 1.5 \\ 
{\tt run5} & 4 & 2.6 & 4 & 1.6 & 3.3 & 12.3 & 0.05 & 0.05 & 0.84 & 12 & 1.4 & 3 & 2.1 & 2.3 & 7.2 & 4.2 & 0.9 & 0.05 & 0.07 & 1.4 \\ 
{\tt run6} & 4 & 3.9 & 6 & 1.6 & 3.3 & 17.3 & 0.05 & 0.05 & 0.84 & 12 & 2.1 & 5 & 2.3 & 2.1 & 7.1 & 4.4 & 1.1 & 0.05 & 0.07 & 1.5 \\ 
{\tt run7} & 4 & 5.3 & 8 & 1.6 & 3.3 & 24.7 & 0.05 & 0.05 & 0.84 & 12 & 2.9 & 7 & 2.4 & 2.2 & 7.0 & 4.4 & 1.2 & 0.05 & 0.07 & 1.5 \\ 
{\tt run8} & 4 & 6.6 & 10 & 1.6 & 3.3 & 30.2 & 0.05 & 0.05 & 0.84 & 12 & 3.6 & 9 & 2.5 & 3.3 & 6.9 & 4.4 & 1.3 & 0.05 & 0.06 & 1.5 \\
\end{longtable}
\end{center}
\end{landscape}
%

\label{lastpage}
\end{document}